# SHAPES AND SPEEDS OF STEADY FORCED PREMIXED FLAMES.

Guy Joulin[1#] and Bruno Denet[2*]

[1] Institut P-prime, UPR 3346 CNRS, ENSMA, Université de Poitiers,
1 rue Clément Ader, B.P. 40109, 86961 Futuroscope Cedex, Poitiers, France.

[2] Aix-Marseille Univ., IRPHE UMR 7342 CNRS, Centrale-Marseille,
Technopole de Château-Gombert, 49 rue Joliot-Curie, 13384 Marseille Cedex 13, France.

**Abstract**.

Steady premixed flames subjected to space-periodic steady forcing are studied *via* inhomogeneous Michelson-Sivashinsky [MS], then Burgers, equations. For both, the flame slope is posited to comprise contributions from complex poles, to locate, and from a base-slope profile chosen in three classes: pairs of cotangents, single-sine functions or sums thereof. Base-slope dependent equations for the pole locations, along with formal expressions for the wrinkling-induced flame-speed increment and the forcing function, are obtained on excluding movable singularities from the latter. Beside exact few-pole cases, integral equations that rule the pole density for large wrinkles are solved analytically. Closed-form flame-slope and forcing-function profiles ensue, along with flame-speed increment *vs.* forcing-intensity curves; numerical checks are provided. The Darrieus-Landau instability mechanism allows MS-flame speeds to initially grow with forcing intensity much faster than those of identically-forced Burgers fronts; only the fractional difference in speed increments slowly decays at intense forcing, which numerical (spectral) time-wise integrations also confirm. Generalizations and open problems are evoked.

## I. INTRODUCTION

The theory of ''thin combustion waves propagating by conduction and chemistry with respect to premixed reactive gases''– *premixed flames*, for short – is delicate: these indeed are free wrinkled boundaries hydrodynamically coupled to adjacent fluids of unequal densities, with vorticity on the burnt side and often vortical forcing motions in the fresh gas. Exactly solvable theoretical models thereof, if basic but encompassing all those aspects, would be invaluable.

Sivashinsky [1] tailored the first evolution equation to describe the dynamics of *unforced* premixed flames; beside curvature effects and an eikonal-type nonlinearity, it accounts for the Darrieus [2]-Landau[3] [DL] hydrodynamic instability for small Atwood numbers $\mathcal{A}$ based on the fresh-to-burnt gas density ratio $E > 1$, $\mathcal{A} := (E-1)/(E+1)$ [definitions are henceforth denoted by " := "]. Once suitably rescaled the nonlinear integro-differential equation so derived in [1], and first studied numerically in [4], reads:

$$\varphi_t + \tfrac{1}{2}(\varphi_x)^2 - \nu\varphi_{xx} + \mathcal{H}[\varphi_x] = 0, \qquad (1.1)$$

and is often called the Michelson-Sivashinsky [MS] equation. The unknown $\varphi(t,x)$ therein represents the flame location and shape when observed in a frame $(x, y)$ that drifts at flat-flame speed $u_L$ towards $y<0$, with $y>\varphi(t,x)$ belonging to the burnt gas. The subscripts stand for partial derivatives in scaled time $t$ and abscissa $x$. The constant $\nu > 0$ controls how the local speed $u_n$ [relative to the fresh mixture] of a front element responds to curvature [5], $1 - u_n/u_L \sim \mathcal{A}^2 \nu \varphi_{xx}$; if the reciprocal pattern wavenumber is used as $x$-wise reference length for periodic $\varphi(t,x)$s, $\nu$ is the neutral-to-actual wavelength ratio. The Hilbert transform $\mathcal{H}[.]$, defined by a principal-value [pv.] integral as $2\pi\mathcal{H}[\varphi_x](x) := \text{pv.} \int_{-\pi}^{+\pi} \varphi_x(x')\cot(\tfrac{1}{2}(x-x'))dx'$, encodes the nonlocal DL instability: since $\mathcal{H}[e^{iqx}] = i.\text{sgn}(q)e^{iqx}$, the growth rate $\varpi$ of normal modes $\varphi \sim e^{iqx+\varpi t}$ of the linearized (1.1) is $\varpi = |q| - \nu q^2$. The nonlinear contribution to $\varphi_t$ in (1.1) is mainly geometric [1]: tilted flame elements propagate along their local normal, at an angle $\gamma(t,x) \sim -\mathcal{A}\varphi_x$ to the mean propagation direction [ $y$-axis], and $1-\cos(\gamma) \sim \mathcal{A}^2 \tfrac{1}{2}(\varphi_x)^2$. Although derived in [1] as a leading order result for $\mathcal{A} \to 0+$, the MS equation still governs spontaneously wrinkled flames when another two orders in $\mathcal{A}$ are retained [6, 7] and the corrections – *e.g.*, vorticity-induced – are absorbed in renormalized coefficients [all scaled out to write (1.1)]; accordingly (1.1) captures [4] the main trends of spontaneous flame evolutions revealed by numerical simulations that employ $\mathcal{A} = O(1)$ [8, 9]. As a matter of fact (1.1) also governs nuclear flames in Ia-Supernovae where $\mathcal{A} = \tfrac{1}{5}$ to $\tfrac{1}{2}$ *is* fairly small [10], and reactive infiltration fronts [11] where a fractional change in porosity plays the part of $\mathcal{A}$.

---

[#] Electronic address: guy.joulin@lcd.ensma.fr ; Corresponding author.
[*] Electronic address: bruno.denet@irphe.univ-mrs.fr .
        

  Exact solutions to (1.1) are accessible in closed form *via* pole-decompositions [12, 13] which convert the evolution of $\varphi(t,x)$ to a many-body problem for complex simple poles $z_k(t)$ of the analytically continued front slope $\varphi_x(t,z)$, $z := x + iB$ [[14] and refs. therein]. The pairwise pole interactions stemming from $(\varphi_x)^2$ in (1.1) make the $z_k(t)$s mutually attract one another in $x$-direction and repel along $B$ [13]; this describes how cusped cells soon form and why the pairwise coalescence of crests [usually sharp, local maxima of $\varphi(t,x)$ ] ultimately builds up a single one in channels endowed with periodic conditions. If $2\pi$-periodicity is assumed, $2N$ poles [per cell] may settle at steady locations $iB_{0,k}$ along the imaginary axis $(\mathrm{mod.}2\pi)$, with $B_{0,-k} = -B_{0,k}$ for the reality of $\varphi(t,x)$ and $N \leq N_{opt}(\nu) := \lfloor \frac{1}{2\nu} + \frac{1}{2} \rfloor$ $[\lfloor . \rfloor :=$ integer part]. Then $\varphi(t,x)$ is $-V_0 t + \varphi_0(x)$, where $\varphi_0(-x) = \varphi_0(x)$ for crests centered on $x = 0$ and [13]:

$$V_0 = \mathcal{V}(\nu N), \quad \mathcal{V}(n) := 2(1-n)n ,$$
$$\varphi_{0x}(x) = -\nu \sum_{k=-N}^{N} \cot(\tfrac{1}{2}(x - iB_{0,k})) . \tag{1.2}$$

The pole altitudes $B_{0,k}$ obey coupled nonlinear *pole-equations*, to be evoked at length later on. Long wrinkles have $N \sim 1/\nu \gg 1$ and densely distributed poles whose density $\rho_0(B)$ obeys a singular integral equation, analytically solved for $N = N_{opt}(\nu) \approx 1/2\nu$ [13] or less [15]: $\rho_0(B)$ vanishes *iff* $1 \geq \tanh(\tfrac{1}{2}|B|) \geq \sin(\pi N \nu)$, and yields the small-$\nu$ slope $\varphi_{0x}(x)$ in closed form. In fair accord [7] with simulations up to $\mathcal{A} = \tfrac{3}{4}$ [$E = 7$] the *non*-scaled, dimensioned effective burning speed $u_{e,0}$ has $u_{e,0}/u_L = 1 + f(E)V_0$, with $f(E) = \tfrac{1}{16}(E - E^{-1})^2/E$ up to $o(\mathcal{A}^4)$ terms.

Such exact results only concern flame propagations into quiescent gases, though. It would be of interest, *per se* and in relation with turbulent flames, to extend this to situations where the fresh gas involves $x$-periodic shear-flows or vortices that will induce extra flame wrinkling. Some works [16, 17] admittedly stressed the need for incorporating the combined influences of DL instability and incoming velocity modulations so as to accurately model weakly- or moderately-turbulent flames, but to date no theoretical work handles the nonlinear forcing/DL interactions *directly*, without using *a priori* averaging [and closures] or numerics [18] [19].

The present contribution aims at reducing the gap. Expectedly, incorporating another cause of wrinkling beside the DL mechanism does not really help one handle the nonlinear and nonlocal Eq.(1.1): its unsteadily forced version has so far been beyond reach of theoretical tools and remains only amenable to numerical ones [20-22]. The present work has a more modest scope: it focuses on *steadily* propagating *forced* wrinkled fronts of permanent shapes, with emphasis on analytical solutions; (1.1) then is the minimally-viable model to start from.

The paper is organized as follows. Section **II** introduces inhomogeneous MS and Burgers equations for steady forced flames, and proposes a method to construct *exact* solutions. This is applied to specific examples in Secs. **III**-**V**. The paper ends up with conclusions, hints of generalizations and open questions [Sec.**VI**]. A technical point is summarized in Appendix **A**.

## II. FORCED PROPAGATIONS

### A. Nonlinear equations for shapes & speeds.

The flames considered here are of permanent shapes $y = \varphi(t,x) = -Vt + \varphi(x)$, where $\varphi(x) = \varphi(x+2\pi) = \varphi(-x)$ is assumed to obey the time-independent, forced version of (1.1):

$$-V + \tfrac{1}{2}(\varphi_x)^2 - \nu\varphi_{xx} + \mathcal{H}[\varphi_x] = u(x). \tag{2.1}$$

The unknown constant *speed increment* $V$ tells how fast the flame drifts as a whole in the moving $(x,y)$ frame of reference; $V$ also measures the excess $\langle \sec(\gamma(x)) - 1 \rangle \sim \mathcal{A}^2 \langle \tfrac{1}{2}(\varphi_x)^2 \rangle$ in front arc-length over that of a flat flame, with $\langle . \rangle := (2\pi)^{-1} \int_{-\pi}^{+\pi} (.)dx$ denoting an $x$-average. The *forcing function* $u(x) = u(x+2\pi) = u(-x)$ has $\langle u \rangle = 0$ and accounts for a periodic shear in the fresh gas, or for variations $1 - u_L(x)/\langle u_L \rangle \sim \mathcal{A}^2 u(x)$ in local flat-flame speed $u_L(x)$ caused by modulated reactant concentrations; $u_{rms} := \langle u^2 \rangle^{1/2}$ will be the forcing *intensity*. One goal here is to relate $V$ and flame shape $\varphi(x)$ to $u_{rms}$ and to the shape $u(x)/u_{rms}$ of $u(x)$; once $V$ is found the forcing-affected effective burning speed $u_e$ has $u_e / u_L = 1 + f(E)V$ and could be plotted *vs.* a more usual forcing intensity $U'/u_L = f(E)u_{rms}$ [$f(E)$ as below (1.2)]. Instability-free fronts $\hat{\varphi}(t,x) = -\hat{V}t + \hat{\varphi}(x)$ will also be studied, *preferably* with the same forcing as in (2.1) for comparisons, by means of the inhomogeneous Burgers equation:

$$-\hat{V} + \tfrac{1}{2}(\hat{\varphi}_x)^2 - \nu\hat{\varphi}_{xx} = \hat{u}(x). \tag{2.2}$$

### B. A semi-inverse strategy.

The Riccati Eq.(2.2) is a disguised Schrödinger equation for $\exp(-\hat{\varphi}/2\nu)$ where $-\hat{u}(x)$ and $\hat{V}$ play the parts of a potential and of energy, respectively. Exact solutions abound, *e.g.* those obtained by the Darboux-Crum transformations [23] that generate new potentials and eigenfunctions from old ones, and can incorporate free parameters to the new $\hat{u}(x)$ for flexibility [23]. Because no Darboux-type trick seems available for the nonlinear *integro*-differential equation (2.1) an elementary – yet possibly related – approach is presented below.

Specifically, the solution $\varphi_x(x)$ to (2.1) is first split as:

$$\varphi_x(x) = \Phi_x(x) + \phi_x(x), \tag{2.3}$$

where the base-slope profile $\Phi_x$ is *chosen* out of suitable classes of meromorphic or entire $2\pi$-periodic functions [specified later] that have $\Phi_x(z) = -\Phi_x(-z)$, are real if $z = x + iB$ is, and are imaginary along $z = iB(\mathrm{mod}.\pi)$.

The unknown component $\phi_x(x) = -\phi_x(-x)$ is *postulated* to have a *pole-decomposition*:

$$\phi_x(x) = -\nu \sum_{k=-N}^{N} \cot(\tfrac{1}{2}(x - iB_k)), \tag{2.4}$$

with the same residues $-2\nu$ as in (1.2) but with new poles $B_k$; $N \geq 0$ is left free, and $\phi_x \equiv 0$ if $N = 0$. An auxiliary function $e(z)$ is next defined in terms of $\phi_x(z)$ and $\Phi_x(z)$ as:

$$e(z) := \tfrac{1}{2}(\phi_x)^2 - \nu\phi_{xx} + \mathcal{H}[\phi_x] + \phi_x \Phi_x + \tfrac{1}{2}(\Phi_x)^2 - \nu\Phi_{xx} + \mathcal{H}[\Phi_x]. \tag{2.5}$$

Plugging (2.4) into (2.5) generically produces a $e(z)$ with singularities wherever $|\phi_x(z)| \to \infty$, but those will also be simple poles at most: thanks to the same residues $-2\nu$ in (2.4) as in (1.2), the $\cot^2(\tfrac{1}{2}(z - iB_k))$ terms from $\tfrac{1}{2}(\phi_x(z))^2$ and $\nu\phi_{xx}(z)$ still mutually annihilate in (2.5).

Next, $\Phi_x(z) - \Phi_x(iB_k) \sim (z - iB_k)$ for $z - iB_k \to 0$ if $\Phi_x(z)$ is analytic at $z = iB_k$ (mod.$2\pi$), hence $\phi_x(z)\Phi_x(z) + \nu \sum_{k=-N}^{k=+N} \Phi_x(iB_k)\cot(\tfrac{1}{2}(z - iB_k))$ also is. The latter weighted cotangents are the only contributions brought about by $\Phi_x \not\equiv 0$ in $e(z)$ that diverge at each $iB_k(\mathrm{mod}.2\pi)$; they will act on a par with like cotangents generated by $\mathcal{H}[\phi_x]$ in (2.5) through:

$$\mathcal{H}[\cot(\tfrac{1}{2}(x - iB_k))] = -1 - i\,\mathrm{sgn}(B_k)\cot(\tfrac{1}{2}(x - iB_k)), \tag{2.6}$$

and with those generated by cross-terms when squaring (2.4), *via* the trigonometric identity

$$\cot(a_k)\cot(a_j) = -1 - \cot(a_k - a_j)[\cot(a_k) - \cot(a_j)] \tag{2.7}$$

particularized to $a_k = \tfrac{1}{2}(z - iB_k)$, $a_j = \tfrac{1}{2}(z - iB_j)$, $j \neq k$.

Besides, a $2\pi$-periodic forcing function $u(x)$ must have singularities somewhere, even if it is the restriction to $z = x$ of a function $u(z)$ analytic in a strip about the real axis. One next requires that the singularities of $u(z)$ be *fixed* ones for $u(x)$ to represent an input; in particular $u(z)$ will necessarily be finite at all the *movable* [*i.e.* solution-dependent] poles $iB_k$ of $\varphi_x(z)$. Since $e(z) \equiv V + u(z)$ for consistency with (2.1), the exclusion of all the $\cot(\tfrac{1}{2}(x - iB_k))$ from $e(z)$ yields $2N$ conditions for $u(z)$ to only have fixed, solution-independent singularities:

$$\nu \sum_{k \neq j,\, j=-N}^{N} \coth(\tfrac{1}{2}(B_k - B_j)) = \mathrm{sgn}(B_k) + i\Phi_x(iB_k). \tag{2.8}$$

Such *pole equations* in principle determine the set $\{B_k\}$ of poles altitudes $B_k$, $k = \pm 1, ..., \pm N$

[*e.g.* $\{B_{0,k}\}$ in (1.2), if $\Phi_x \equiv 0$], anticipated not to coincide with the singularities of $u(z)$. The front shape $\varphi(x)$ ensues from (2.3)(2.4) and elementary integration. One can next deduce the flame-speed increment $V$, and even *define* the forcing function $u(x)$, by:

$$V = \langle e(x) \rangle, \quad u(x) := e(x) - \langle e(x) \rangle. \tag{2.9}$$

In this *semi-inverse* strategy for constructing exact solutions to (2.1) – *viz.*: choose $\Phi_x$, get the $B_k$s, deduce $\varphi_x$, $V$ and $u(x)$ – the chore is mostly transferred to solving (2.8): that is not a trivial task either, but methods tailored for unforced fronts [13] can hopefully be adapted.

It remains to specify base-slopes $\Phi_x(x)$. Beside analytical accessibility $u(x)$ deduced from (2.9) should preferably be flexible enough to represent various conditions of forcing, which is not a given : $u(x)$ can still depend on $\{B_k\}$ once the conditions (2.8) are met. In absence of a systematics to identify *the* good $\Phi_x$s the next sections exploit a few ones selected after trials.

To apply the same machinery to the forced Burgers equation (2.2) one just has to delete all DL contributions [$\mathcal{H}[.]$, sgn(.) functions] and use variables with overhats: $\hat{\Phi}_x, \hat{\phi}_x, \hat{\varphi}_x, \hat{N}, \hat{V}...$ .

### III. COTANGENT BASE-SLOPES

#### A. General setting.

The first base-slope profiles $\Phi_x$ to be presently envisaged read:

$$\Phi_x(x) = -m\nu[\cot(\tfrac{1}{2}(x+i\beta)) + \cot(\tfrac{1}{2}(x-i\beta))], \tag{3.1}$$

where $\beta > 0$ and $m$ [any sign] are control parameters. Except for its residue $-2m\nu$ at the fixed poles $z = \pm i\beta(\mathrm{mod}.2\pi)$ $\Phi_x$ has the same structure as (2.8) itself, which on using (2.7) eases the computation of $\phi_x(z)\Phi_x(z)$ in (2.5) and of $u(x)$ in (2.9); this also allows for checks.

The pole equations to consider here follow from specializing (2.8) to (3.1):

$$\nu \sum_{k \neq j=-N}^{+N} \coth(\tfrac{1}{2}(B_k - B_j)) = \mathrm{sgn}(B_k) + m\nu[\coth(\tfrac{1}{2}(\beta - B_k)) - \coth(\tfrac{1}{2}(\beta + B_k))]. \tag{3.2}$$

Their Burgers counterparts involve $\{\hat{B}_k\}, \hat{N}, \hat{m}, \hat{\beta}$ and have no DL-related sgn(.) function.

To obtain $u(x)$ and $V$ one substitutes (3.1) in (2.5) and uses (3.2)(2.9), to ultimately produce:

$$\begin{aligned} u(x) = {} & 2m(m-1)\nu^2 \partial h(x,\beta)/\partial\beta + \\ & 2m\nu[S_N(\beta,\{B_k\}) + m\nu K - 1]h(x,\beta), \end{aligned} \tag{3.3}$$

$$V = \mathcal{V}(\nu(N+m)) + 2m\nu[S_N(\beta,\{B_k\}) + m\nu K - 1], \tag{3.4}$$

with $K := \coth(\beta)$ and the same $\mathcal{V}(n) = 2(1-n)n$ as in (1.2); the auxiliary functions $h(x, \beta)$ and $S_N(\beta, \{B_k\})$ are defined as:

$$h(x, \beta) := \frac{\sinh(\beta)}{\cosh(\beta) - \cos(x)} - 1 = 2\sum_{q=1}^{\infty} e^{-q\beta} \cos(qx), \quad (3.5)$$

$$S_N(\beta, \{B_k\}) := \nu \sum_{k=-N}^{N} \coth(\tfrac{1}{2}(\beta + B_k)). \quad (3.6)$$

If $m = 0$, $u(x)$ vanishes and $V$ resumes the value $V_0 = \mathcal{V}(\nu N)$ [see (1.2)], as expected. $u(x)$ also disappears for $m = 1$ if $S_N(\beta, \{B_k\}) + \nu \coth(\beta) = 1$: that also is (1.2) once written for unforced flames with $N+1$ poles pairs and $B_{0,N+1}$ renamed as $\beta$; $V$ is then $\mathcal{V}(\nu(N+1))$.

As warned in Sec. **II**, $u(x)$ in (3.3) contains a term $S_N(\beta, \{B_k\})$ still involving pole altitudes that themselves depend on $m, \nu, \beta$ and $N$ via (3.2); by (3.3) the shape of $u(x)$ will vary as $u_{rms}$ does at fixed $(N, \nu, \beta)$, which affects the response of $V$ to $u_{rms}$ [see **III**.B4 for a cure].

## B. Large wrinkles.

Following Sec. **II** one must first solve (3.2) to obtain the set $\{B_k\}$ for selected $m, \nu, \beta, N$; (3.4)(3.6) next give the speed increment $V$, and the forcing function $u(x)$ follows from (3.3)-(3.6). The situations with $N = 0$ [no pole at all, $\varphi(x) \equiv \Phi(x)$] or $N = 1$ are the simplest, but not to overload Sec. **III** a discussion of similar particular cases is deferred until Sec. **IV**.

### 1. Density and slope profiles.

Solving (3.2) generically requires numerical methods but the situation simplifies in the limit $\nu \to 0^+$, with $N \sim 1/\nu$ as for unforced flames [13,15] and with $m \sim 1/\nu$ here. The solutions to (3.2) get describable by such a *pole density* $\rho(B, \beta) = \rho(-B, \beta)$ that $\rho(B, \beta)dB$ is the number of $B_k$s located in $[B, B + dB]$. And (3.2) itself becomes an integral equation for $\rho(B, \beta)$ involving a principal-value [pv.] integral, as to cope with the constraint $j \neq k$ in (3.2), and the *known* functions $f_0(B) := \frac{1}{\nu} \text{sgn}(B)$, $f_\star(B, \beta) := m[\coth(\tfrac{1}{2}(\beta - B)) - \coth(\tfrac{1}{2}(\beta + B))]$:

$$\text{pv.} \int_{-B_{\max}}^{+B_{\max}} \frac{\rho(B', \beta)dB'}{\tanh(\tfrac{1}{2}(B - B'))} = f_0(B) + f_\star(B, \beta). \quad (3.7)$$

Implicit in (3.7) is the assumption that $\rho(|B| > B_{\max}, \beta) = 0$ for some $B_{\max} > 0$; also, to get $u(x) \to 0$ for $\beta \to \infty$ at fixed $\nu N$, only densities with $\rho(|B| < B_{\max}, \beta) > 0$ are retained.

To solve (3.7) the independent variable $B$ is first traded for $T := \tanh(\tfrac{1}{2}B)$, which gives:

$$\text{pv.} \int_{-T_{\max}}^{+T_{\max}} \frac{\rho(T',\beta)dT'}{T-T'} = F_0(T) + F_\star(T,\beta), \tag{3.8}$$

$$F_0(T) = \frac{\text{sgn}(T)}{2\nu}, \quad F_\star(T,\beta) = m\frac{(1-\Theta^2)T}{\Theta^2-T^2}, \tag{3.9}$$

where $T_{\max} := \tanh(\frac{1}{2}B_{\max})$, $\Theta := \tanh(\frac{1}{2}\beta)$ and $\rho(T,\beta)$ is shorthand for $\rho(B,\beta)|_{B=2\tanh^{-1}(T)}$. Equation (3.8) is of the Tricomi type [24]: its solution could be written as a singular integral [25] but here it is simpler to exploit its *linearity*. Let $\rho_0(T)$ [or $\rho_\star(T,\beta)$] be the solution, vanishing at $|T|=T_{\max}^-$, to (3.8) when the right-hand side reduces to $F_0(T)$ [or $F_\star(T,\beta)$)]. $\rho_0(T)$ was already computed for unforced flames [15]; up to a constant factor, the partial problem for $\rho_\star(T,\beta)$ happens to also rule the density of zeros of Gegenbauer polynomials $\mathcal{C}_{2N}^{(\lambda)}(T/\Theta)$ at $\lambda \sim N \gg 1$ [26-28] provided $2N/\lambda = (1-T_{\max}^2/\Theta^2)^{-1/2}-1$. One ultimately gets:

$$\rho_0(T) = \frac{1}{\nu\pi^2}\cosh^{-1}(\frac{T_{\max}}{|T|}), \tag{3.10}$$

$$\rho_\star(T,\beta) = \frac{m(1-\Theta^2)}{\pi(1-T_{\max}^2/\Theta^2)^{1/2}} \frac{(T_{\max}^2-T^2)^{1/2}}{(\Theta^2-T^2)}, \tag{3.11}$$

and the linearity of (3.8) yields the pole density at once: $\rho(T,\beta) = \rho_0(T) + \rho_\star(T,\beta)$.

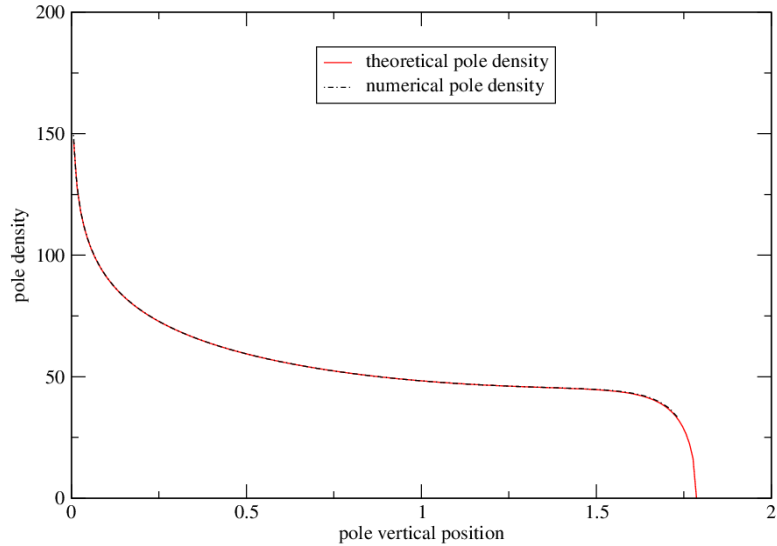

FIG.1. (Color online): Theoretical pole density $\rho(B,\beta)$ from Eqs.(3.10)(3.11) (red solid line), and numerical density $\rho_{num}(B)$ deduced from Newton resolution of (3.2) (black dash-dot line), *vs.* pole altitude $B$; both have $N=100$, $\nu=1/199.5$, $m=50$ and $\beta=2$.

To complete $\rho(T,\beta)$, the needed $T_{max} = \tanh(\tfrac{1}{2}B_{max})$ follows from density normalization:

$$N = \int_0^{B_{max}} \rho(B,\beta)dB = \frac{\sin^{-1}(T_{max})}{\pi\nu} + m\left[\frac{(1-T_{max}^2)^{1/2}}{(1-T_{max}^2/\Theta^2)^{1/2}} - 1\right]. \quad (3.12)$$

As happened with stretched flames [29] the above $\rho_0(T) + \rho_*(T,\beta)$ can get locally negative near $|T| = T_{max}$, and is then *spurious* as density profile. Expanding (3.10)(3.11) at $|T| \lesssim T_{max}$ reveals that the genuine, nonnegative densities $\rho(T,\beta)$ must satisfy:

$$(1-\Theta^2)m\nu \geq -\frac{(\Theta^2 - T_{max}^2)^{3/2}}{\pi \Theta T_{max}}, \quad (3.13)$$

which can possibly be violated if $m < 0$: poles then likely detach off the main pile [29].

Figure 1 compares (3.10)-(3.12) with a numerical density $\rho_{num}(B)$ evaluated from Newton resolutions of (3.2) as $\rho_{num}(\tfrac{1}{2}(B_k + B_{k-1})) = 1/(B_k - B_{k-1})$, for $N = 100$, $1/\nu = 199.5$, $\beta = 2$, $m = 50$: good accord is obtained wherever $\rho_{num}(.)$ is available, *i.e.* for $|B| \leq \tfrac{1}{2}(B_N + B_{N-1})$.

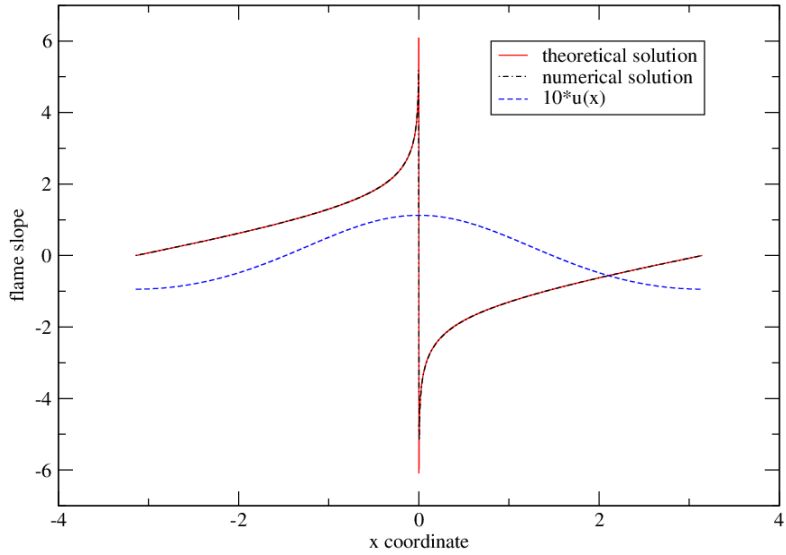

FIG.2. (Color online): Theoretical MS front slope $\varphi_x(x) = \varphi_{0x}(x) + \varphi_{*x}(x,\beta)$ from (3.14)(3.15) (red solid line) and numerical slope (black dash-dot line) deduced from (2.3)(3.1) and Newton resolution of (3.2), *vs.* abscissa $-\pi \leq x \leq \pi$; same parameters as in Fig.1. The dashed blue line is $10 \times u(x)$ deduced from (3.18) and (3.5)(3.17).

The above density $\rho(T,\beta)$ gives access to the $N \sim 1/\nu \gg 1$ version of (2.4), viz. $\phi_x(x) = -\nu\int_{-B_{max}}^{+B_{max}} \rho(B',\beta)\cot(\tfrac{1}{2}(x-iB'))dB'$, which on analytical evaluation from (3.10)(3.11) yields the 'total' MS front slope $\varphi_x(x) = \Phi_x(x) + \phi_x(x)$ in the closed form $\varphi_{0x}(x) + \varphi_{*x}(x,\beta)$, where:

$$\varphi_{0x}(x) = -\frac{2}{\pi}\sinh^{-1}(\frac{T_{max}}{\vartheta(x)}), \quad \vartheta(x) := \tan(\tfrac{1}{2}x), \tag{3.14}$$

$$\varphi_{\star x}(x,\beta) = -\frac{2m\nu(1-\Theta^2)}{(1-T_{max}^2/\Theta^2)^{1/2}} \frac{\vartheta(x)(1+T_{max}^2/\vartheta(x)^2)^{1/2}}{(\Theta^2+\vartheta(x)^2)} \quad . \tag{3.15}$$

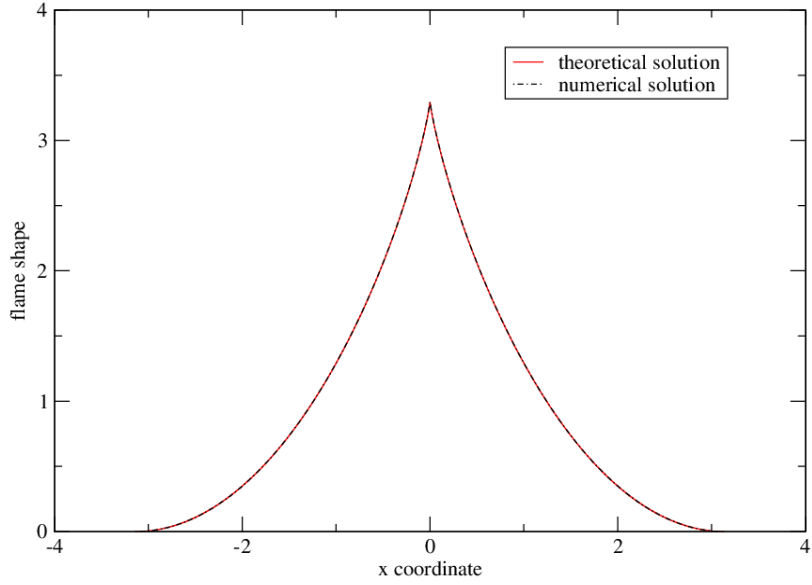

FIG.3. (Color online): Theoretical MS front shape $\varphi(x)=\varphi_0(x)+\varphi_\star(x,\beta)$ (red solid line) corresponding to Figs. 1 & 2, and numerical one (black dash-dot line), both *vs.* $-\pi \leq x \leq \pi$. The unforced front ($m=0$) with $N\nu = 100/199.5$ has $\varphi(0)-\varphi(\pm\pi) \approx 2.33$ [15].

Interestingly, the base-slope $\Phi_x(x)$ does not show up additively in $\varphi_x(x)$. The wavelength ratio $\nu \ll 1$ affects $\varphi_x(x)$ only through $m\nu = O(1)$ and $N\nu = O(1)$, *via* (3.12)(3.15), whence (3.14)(3.15) yield $\varphi(0)-\varphi(\pi) = O(1)$: long wrinkles also are large; all have a cusp at $x=0$ due to the DL contribution $\varphi_0(x)$, $\varphi(x)-\varphi(0) \sim |x|\ln(|x|)$. Figures 2,3 compare predicted and numerical $\varphi_x(x)$ and $\varphi(x)$; the associated $u(x)$, from (3.18) below, also is displayed in Fig.2.

The Burgers fronts formally have $\hat{\rho}_0(B) \equiv 0$: $\hat{\rho}(T,\hat{\beta}) \equiv \hat{\rho}_\star(T,\hat{\beta})$ and $\hat{\varphi}_x(x) \equiv \hat{\varphi}_{\star x}(x,\hat{\beta})$ are still given by (3.11)(3.15), yet with parameters $\hat{m}$, $\hat{\Theta} = \tanh(\tfrac{1}{2}\hat{\beta})$ and $\hat{T}_{max} := \tanh(\tfrac{1}{2}\hat{B}_{max})$ in lieu of $m, \Theta$ and $T_{max}$. A normalization that parallels (3.12) gives $\hat{T}_{max}$ itself, *viz.*:

$$\hat{N} = \hat{m}[\frac{(1-\hat{T}_{max}^2)^{1/2}}{(1-\hat{T}_{max}^2/\hat{\Theta}^2)^{1/2}} - 1] \quad . \tag{3.16}$$

Necessarily $\hat{m} > 0$, as is patent in the Burgers version of (3.2) written for $\hat{B}_{\hat{N}} = \max_k(\hat{B}_k) < \hat{\beta}$.

Figure 4 compares a predicted DL-free front slope $\hat{\varphi}_x(x)$ with what Newton iterations give from the analog of (3.2); the associated $\hat{u}(x)$, *e.g.* deduced from (3.22) below, is also shown.

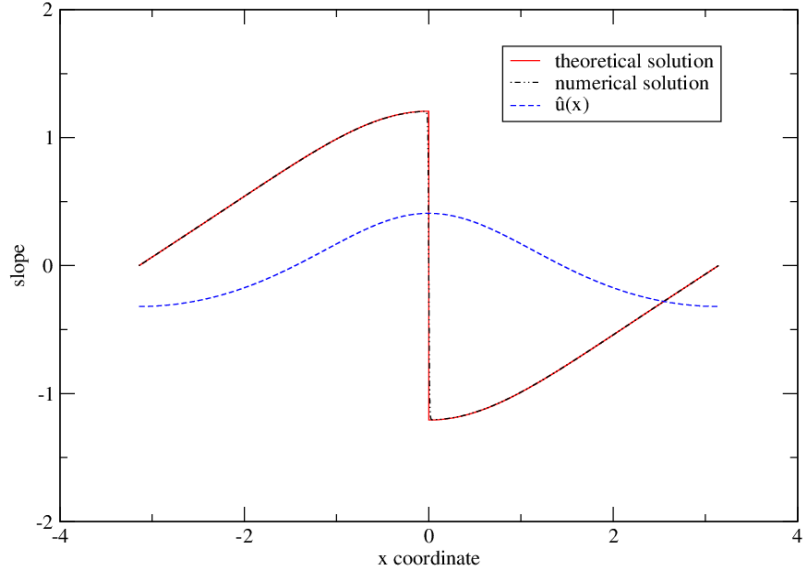

FIG.4. (Color online): Front slope $\hat{\varphi}_{*x}(x,\hat{\beta})$ from the Burgers version of (3.15) (3.16) (red solid line) for $\hat{\beta}=2, \hat{N}=100, \nu=1/199.5, \hat{m}=50$, and numerical one (black dash-dot line), *vs.* $-\pi \le x \le \pi$. The dashed blue line is $\hat{u}(x)$, from (3.22) and the analogs of (3.15)(3.20).

## 2. *Forcing function and speed increment.*

According to Sec. **II**, the next steps are to compute the flame-speed increment $V$ and the forcing function $u(x)$ belonging to long MS wrinkles. The sum $S_N(\beta,\{B_k\})$ in (3.6) is first replaced by $S_\infty(\beta,T_{max}) := \nu \int_{-B_{max}}^{+B_{max}} \coth(\tfrac{1}{2}(\beta+B))\rho(B,\beta)dB$; thanks to (3.10)(3.11) this is also:

$$S_\infty(\beta,T_{max}) = \frac{2}{\pi}\sin^{-1}(\frac{T_{max}}{\Theta}) + \frac{m\nu(1-\Theta^2)}{\Theta}\frac{T_{max}^2}{\Theta^2-T_{max}^2} \ . \qquad (3.17)$$

The ensuing expressions of $u(x)$, $u_{rms}^2$ [deduced by plugging the series (3.5) in (3.3), squaring, averaging and then resuming the result] and of $V$ are:

$$\begin{aligned}u(x) = {}& 2m^2\nu^2 \partial h(x,\beta)/\partial\beta + \\ & 2m\nu[S_\infty(\beta,T_{max})+m\nu K-1]h(x,\beta) \ ,\end{aligned} \qquad (3.18)$$

$$\begin{aligned}u_{rms}^2 = {}& 4m^2\nu^2[(S_\infty(\beta,T_{max})+m\nu K-1)]^2(K-1) + \\ & 2m^4\nu^4 K(K^2-1) - 4m^3\nu^3[S_\infty(\beta,T_{max})+m\nu K-1](K^2-1) \ ,\end{aligned} \qquad (3.19)$$

$$V = \mathcal{V}((N+m)\nu) + 2m\nu[S_\infty(\beta,T_{max})+m\nu K-1] \ . \qquad (3.20)$$

Figure 5 displays $V$ vs. $u_{rms}$ curves plotted in parametric form [with $T_{max}$ as running variable] at sample *fixed* $\nu N$ s and for $\beta = 2$: given a current $T_{max}$, (3.12) yields $mv$ to access the corresponding pole-density and front-slope profiles from (3.10)(3.11) and (3.14)(3.15), respectively, *iff* (3.13) is met; $u(x)$ and then $(u_{rms}, V)$ belonging to *this* $T_{max}$ [only] ensue from (3.17)-(3.20). All such $V$ vs. $u_{rms}$ response curves have a branch pertaining to in-phase fronts, where the DL effect and forcing cooperate to push the crests towards $\varphi(x) > 0$ [$u(0) > 0$, $-\mathcal{H}[\varphi_x](0) > 0$]; in accord with the estimates given in subsection **III**.B3, $V \sim u_{rms}$ if $u_{rms} \gg 1$.

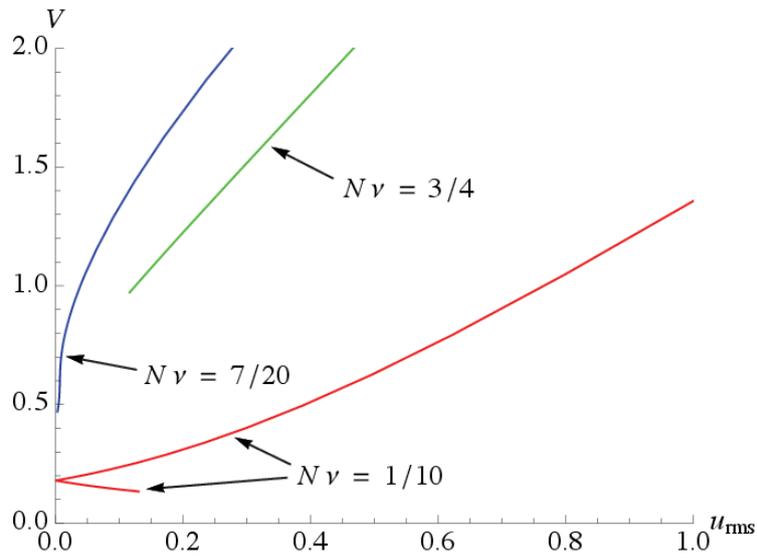

FIG.5. (Color online): Speed increment $V$ vs. forcing intensity $u_{rms}$ for MS flames from (3.20)(3.19) for $\beta = 2$, $N\nu = \frac{1}{10}$ (red), $N\nu = \frac{7}{20}$ (blue) or $N\nu = \frac{3}{4}$ (green). The semi-infinite branches have $V \sim u_{rms}$ at $u_{rms} \gg 1$; the upmost ones have $V \geq \mathcal{V}(N\nu) + \frac{2}{\Theta}(N\nu - \frac{1}{\pi}\sin^{-1}(\Theta))^2$ and $u_{rms}^2 \geq \frac{2}{\Theta^3}(1-\Theta)^2(N\nu - \frac{1}{\pi}\sin^{-1}(\Theta))^4$, here with $\Theta = \tanh(1) = 0.7616$ and $\frac{1}{\pi}\sin^{-1}(\Theta) = 0.2756$.

If $\nu N > \frac{1}{\pi}\sin^{-1}(\Theta)$, $u_{rms} = 0$ is *not* reached: poles pile up at $|B| \lesssim \beta$ and $\varphi_{\star x}(x, \beta)$ in (3.15) does not vanish for $T_{max} \to \Theta^-$ [$mv \to 0$ then, but $mv/(\Theta^2 - T_{max}^2)^{1/2}$ stays finite by (3.12)]. When $\nu N < \frac{1}{\pi}\sin^{-1}(\Theta)$ though, the semi-infinite $V(u_{rms})$ branch touches the $V$-axis at a point where $T_{max} = \sin(\pi\nu N) < \Theta$, $mv = 0$, $u_{rms} = 0$, and $V_{u\equiv 0} = \mathcal{V}(\nu N) < \frac{1}{2}$ by (3.20). Another branch with $V \leq V_{u\equiv 0}$ then also exists, Fig.5, and pertains to fronts in phase-opposition with $u(x)$ for which forcing hinders the DL effect [$-\mathcal{H}[\varphi_x](0) > 0$, $u(0) < 0$]; the lower branch is cut beyond a cusp singularity [$\partial u_{rms}/\partial T_{max} = 0 = \partial V/\partial T_{max}$] at which (3.13) starts failing.

## 3. MS vs. Burgers.

The above material is adapted to Burgers fronts with a few changes: $m$, $K$ and $m\nu K - 1$ are replaced by $\hat{m}$, $\hat{K} := \coth(\hat{\beta})$ and $\hat{m}\nu\hat{K}$ in (3.18)-(3.20), respectively, $\mathcal{V}(\nu(N+m))$ needs to be modified to $-2(\nu(\hat{N}+\hat{m}))^2$ instead of (1.2), and $S_\infty(\beta, T_{max})$ of (3.17) is replaced by:

$$\hat{S}_\infty(\hat{\beta}, \hat{T}_{max}) = \frac{\hat{m}\nu(1-\hat{\Theta}^2)}{\hat{\Theta}} \frac{\hat{T}_{max}^2}{\hat{\Theta}^2 - \hat{T}_{max}^2} . \tag{3.21}$$

The forcing function $\hat{u}(x)$ obtained as indicated above will generically differ from $u(x)$. For MS fronts [from (2.1)] and Burgers counterparts [from (2.2)] to be forced by the *same* $\hat{u}(x) \equiv u(x)$, (3.18) and (3.5) require $\hat{\beta} = \beta$, $\hat{m} = m$ and then $\hat{S}_\infty(\beta, \hat{T}_{max}) = S_\infty(\beta, T_{max}) - 1$. The latter condition yields $\hat{T}_{max} = \tanh(\frac{1}{2}\hat{B}_{max})$ [needed in $\hat{\rho}(T, \beta)$ and $\hat{\varphi}_x(x)$] in terms of $T_{max}$, itself accessible from (3.12) as a function of the control parameters $\beta$, $N\nu$, $m\nu$. By (3.16), $\hat{N}$ must also be adjusted for overall consistency, which for a generic $\nu = O(1)$ would have not been feasible with an integer $\hat{N}$. But here $\nu = 0^+$ and only the *real* grouping $\hat{N}\nu$ matters, for use in the *leading*-order Burgers density $\hat{\rho}_*(T, \beta)$, front slope $\hat{\varphi}_{*x}(x)$ and speed increment $\hat{V}$.

Before doing so, however, an important remark is due: for long wrinkles $\hat{\varphi}_x(x)$ and $\hat{V}$ can be determined in an 'easy way', much simpler than just described, as soon as $\hat{u}(x)$ is available, be it equal to $u(x)$ or not. From the $\nu = 0^+$ version $\frac{1}{2}(\hat{\varphi}_x)^2 = \hat{u}(x) + \hat{V}$ of (2.2) one indeed has:

$$\hat{V} + \hat{u}_{min} = 0, \quad \tfrac{1}{2}(\hat{\varphi}_x)^2 = \hat{u}(x) - \hat{u}_{min} , \tag{3.22}$$

provided $\hat{u}(x)$ has its *global* minimum $\hat{u}_{min}$ at $x = \hat{x}_{min}$ and $\hat{\varphi}_x(\hat{x}_{min}) = 0$. Though not manifest at first glance, $\hat{u}(x) + \hat{V}$ and $\hat{\varphi}_{*x}(x)$ as given by the Burgers counterparts of (3.18)(3.20) and (3.15) do obey (3.22), provided (3.16) holds. Across a crest, say sat at $x = x_c$, $\hat{\varphi}_x(x_c^+) = -\hat{\varphi}_x(x_c^-) \neq 0$ [see Fig.4] and only $\hat{\varphi}_x(x_c^+) < 0$ matches the inner crest structure at $x - x_c \sim \nu$; this *in fine* dictates the sign of $\hat{\varphi}_{xx}(\hat{x}_{min}) = +(\hat{u}_{xx}(\hat{x}_{min}))^{1/2}$: $\hat{\varphi}(x)$ has troughs at $\hat{x}_{min} (\mathrm{mod}. 2\pi)$.

If applied to the MS equation the above reasoning yields $V + u_{min} = \mathcal{H}[\varphi_x](x_{min})$ at $\nu = 0^+$. As (2.4)(2.6) imply $\mathcal{H}[\varphi_x](\pi) > 0$, MS *flames are faster than* Burgers *ones when* $\hat{u}(x) = u(x)$ [*i.e.* $V > \hat{V}$], *if* both fronts have their troughs at $\hat{x}_{min} = x_{min} = \pi (\mathrm{mod}. 2\pi)$; the latter 'if' is to stress that, although Burgers fronts with a crest at $x=0$ and a trough at $x=\pi$ cannot resist $\hat{m}<0$, crest and trough may then swap their locations provided this leads to $\hat{u}_{\min}=\hat{u}(0)$.

From (3.22) $(\hat{\varphi}_x)^2$ is expected to scale like $\hat{u}_{\min}$, in which case $\hat{V}\sim\hat{u}_{rms}$ at intense forcing if $\hat{u}(x)/\hat{u}_{rms}$ stays $O(1)$. Likewise, $\varphi_x\sim u_{rms}^{1/2}$ results in $V+u_{\min}=\mathcal{H}[\varphi_x](x_{\min})\sim u_{rms}^{1/2}$ for those MS fronts that survive $u_{rms}\gg 1$, implying $V-\hat{V}\sim +u_{rms}^{1/2}$ when $\hat{u}(x)=u(x)\gg 1$: only the *fractional* correction $V/\hat{V}-1$ to the Burgers speed increment caused by the DL effect then disappears, yet only slowly. Not all MS fronts survive intense forcing, though, see Fig.5.

## 4. Fixed-shape forcing.

When $u(x)$ is given by (3.5)(3.17)(3.18) its shape varies with the intensity $u_{rms}$ at fixed $(N\nu,\beta)$, which $V$ 'feels'. Choosing $A:=4m\nu e^{-\beta}$, $\alpha:=(S_\infty(\beta)-1)/m\nu+K$ and $\beta$ as *new control parameters* eliminates this bias. For, the forcing function in (3.18) also is:

$$u(x)=\tfrac{1}{8}A^2(\frac{\partial}{\partial\beta}+\alpha)h(x,\beta)=\tfrac{1}{4}A^2 e^{2\beta}\sum_{q=1}^{\infty}(\alpha-q)e^{-q\beta}\cos(qx),\qquad(3.23)$$

whereby $(\alpha,\beta)$ control its shape and *only* $A$ does its intensity. Equation (3.17) relates $T_{\max}=\tanh(\tfrac{1}{2}B_{\max})$ and $(A,\alpha,\beta)$, and the normalization (3.12) links $N$ to all:

$$\frac{Ae^\beta}{4}[\frac{(1-\Theta^2)}{\Theta}\frac{T_{\max}^2}{\Theta^2-T_{\max}^2}+K-\alpha]=\frac{2}{\pi}\cos^{-1}(\frac{T_{\max}}{\Theta}),\qquad(3.24)$$

$$2\nu N=\frac{Ae^\beta}{2}[\frac{(1-T_{\max}^2)^{1/2}}{(1-T_{\max}^2/\Theta^2)^{1/2}}-1]+\frac{2}{\pi}\sin^{-1}(T_{\max}).\qquad(3.25)$$

The total pole weight $N\nu$ is now *determined* by $u_{rms}\sim A^2,(\alpha,\beta)$; and, importantly, $\nu N_{u\equiv 0}=\tfrac{1}{\pi}\sin^{-1}(\Theta)$. The speed increment $V$ from (3.20) and the intensity $u_{rms}$ in (3.19) are recast as:

$$\begin{aligned}V&=\mathcal{V}(N\nu)+\tfrac{1}{2}Ae^\beta[\tfrac{1}{4}Ae^\beta(\alpha-1)+1-2N\nu]\ ,\\ u_{rms}^2&=\tfrac{1}{128}A^4 e^{4\beta}(K-1)[2\alpha^2+(K+1)(K-2\alpha)]\ .\end{aligned}\qquad(3.26)$$

At fixed $(\alpha,\beta)$, $u_{rms}$ and the minimum value $u_{\min}$ of $u(x)$ thus are merely proportional to $A^2$. By (3.24), $A\sim(\Theta^2-T_{\max}^2)^{3/2}$ if $T_{\max}\to\Theta^-$: $u_{rms}=0$ is now reached, (3.25) does give $2\nu N_{u\equiv 0}=\tfrac{2}{\pi}\sin^{-1}(\Theta)$, and $V_{u\equiv 0}=\mathcal{V}(\tfrac{1}{\pi}\sin^{-1}(\Theta))=\tfrac{1}{2}-O(e^{-\beta})$. Next, $V-V_{u\equiv 0}$ grows as $\nu N-\tfrac{1}{\pi}\sin^{-1}(\Theta)\sim A^{2/3}\sim u_{rms}^{1/3}$ for $u_{rms}\ll 1$, similar to what simulations gave [22, 30]. Such a rapidly rising initial $V$ is also compatible with experiments [31] at high pressures [whence $\nu\ll 1$], as is the prediction $V(u_{rms} \gg 1) \sim A^2 \sim u_{rms}$ of (3.24)-(3.26) for $\alpha \geq K$; the latter limit corresponds to $T_{\max}^2$ approaching the value $0 \leq T_{crit}^2 < \Theta^2$ that makes the factor of $A$ in (3.24) vanish. If $\alpha < K$ the whole range $T_{\max}^2 < \Theta^2$ is allowed, $\partial N / \partial u_{rms}\big|_\alpha < 0$ and each $V$ vs. $u_{rms}$ curve *ends* at *finite* $V$ and $u_{rms}$, $\frac{1}{4} A e^\beta (K - \alpha) = 1$ and $T_{\max} = 0^+ = N\nu$: at the end-point, $\varphi_x(x) = \Phi_x(x)$. The resulting array of $V$ vs. $u_{rms}$ curves belonging to sample fixed $\alpha \lessgtr K$ [hence to fixed $u(x)$ shapes] is very similar to Fig.8 in the next Section and is omitted.

When $\hat{u}(x) = u(x)$, the Burgers speed-increment $\hat{V}$ predicted by (3.5)(3.23) and (3.22) is far simpler: at fixed $(\alpha, \beta)$, $\hat{V} = -u_{\min} \propto u_{rms}$ for all $u_{rms}$ [fixed $\nu N$ s only gave $\hat{V} = O(u_{rms})$ ].

## IV. SIMPLER BASE-SLOPE

The simplest base-slope is $\Phi_x(x) = -A_1 \sin(x)$, $A_1 = const.$, and 'its' pole equations (2.8) are:

$$\nu \sum_{k \neq j = -N}^{+N} \coth(\tfrac{1}{2}(B_k - B_j)) = \text{sgn}(B_k) + A_1 \sinh(B_k) . \tag{4.1}$$

Interestingly, (4.1) also follows from (3.2) in a suitable *double limit*: $\beta \to \infty$, $m\nu \to \infty$ with $4m\nu e^{-\beta} := A_1$ kept fixed. Two ways to $\varphi_x$, $V$, $u(x)$ thus exist : redo all the steps defined in Sec. **II.B**, or employ the double limit to simplify the findings of Sec. **III.A**. Both ways give:

$$u(x) = (\nu(2N+1) - 1) A_1 \cos(x) - \tfrac{1}{4} A_1^2 \cos(2x) ,$$
$$u_{rms}^2 = \tfrac{1}{2}(\nu(2N+1) - 1)^2 A_1^2 + \tfrac{1}{32} A_1^4 , \tag{4.2}$$

$$V = \mathcal{V}(\nu N) + \tfrac{1}{4} A_1^2 + 2 A_1 \nu \sum_{k=1}^{N} \cosh(B_k) , \tag{4.3}$$

with the same $\mathcal{V}(\nu N) = 2(1 - \nu N)\nu N$ as in (1.2). Remarkably, the new forcing function $u(x)$ is *independent of the pole locations*. It is also invariant by $(A_1, x) \leftrightarrow (-A_1, x - \pi)$ at fixed $(N, \nu)$, which originates from the fact that (4.1)-(4.3) also are the double limit of a variant of Sec. **III** where the selected base-slope profile differs from (3.1) by $(m, \beta) \to (-m, \beta + i\pi)$, $\Phi_x(x) \to m\nu[\cot(\tfrac{1}{2}(x - \pi + i\beta)) + \cot(\tfrac{1}{2}(x - \pi - i\beta))]$; this leaves $u(x)$ almost unchanged if $\beta \geq 2$ [ $x \to x - \pi$ nearly compensates $m \to -m$ ] and produces $V$ vs. $u_{rms}$ curves similar to Fig.5.

Burgers fronts have no $\text{sgn}(.)$ function in (4.1), $\nu(2N+1) - 1$ is replaced by $\nu(2\hat{N}+1)$ in (4.2) and $\mathcal{V}(\nu N)$ by $-2(\nu\hat{N})^2$ in (4.3) .

## A. Few-pole fronts

The 'poleless' MS flame with $N=0$, $\phi(x) \equiv 0$ and $\varphi(x) = A_1 \cos(x)$ is the simplest: (4.1) is then void, $u(x) = (\nu-1)A_1 \cos(x) - \frac{1}{4} A_1^2 \cos(2x)$, $u_{rms}^2 = \frac{1}{2}(\nu-1)^2 A_1^2 + \frac{1}{32} A_1^4$, and $V = \frac{1}{4} A_1^2$ is $\sim u_{rms}^2$ if small or $\sim u_{rms}$ when large; the Burgers analog has $\hat{V} = \frac{1}{4} \hat{A}_1^2$, and $\nu$ instead of $\nu-1$. The signs of $\hat{A}_1, A_1$ do not matter here, for changing one amounts to a global shift $x \to x-\pi$. No restriction on $\nu$ exists for $\varphi(x)$ or $\hat{\varphi}(x)$ to *separately* read $A_1 \cos(x)$ or $\hat{A}_1 \cos(x)$. But requiring $\hat{u}(x) = u(x)$ leads to $\hat{A}_1 = \pm A_1$, $\hat{V} = V$, $\nu = \pm(\nu-1)$, which is viable *iff* $\hat{A}_1 = -A_1$ and $\nu = 1-\nu = \frac{1}{2}$; when $\hat{u}(x) = u(x)$, $\hat{\varphi}(x)$ cannot be a mere $\hat{A}_1 \cos(x)$ whenever $\nu \neq \frac{1}{2}$.

The $N=1$ case also is simple, for (4.1) yields $A_1 = (\nu \coth(B_1) - 1)/\sinh(B_1)$, (4.2) becomes $u(x) = (3\nu-1)A_1 \cos(x) - \frac{1}{4} A_1^2 \cos(2x)$ and the curve $V$ vs. $u_{rms}$ is accessible in parametric form from (4.2)(4.3); and similarly, but separately, for Burgers fronts where $\nu \coth(\hat{B}_1)/\sinh(\hat{B}_1) = \hat{A}_1$ and $\hat{u}(x) = 3\nu \hat{A}_1 \cos(x) - \frac{1}{4} \hat{A}_1^2 \cos(2x)$ if $\hat{N} = 1$. However, imposing $\hat{u}(x) = u(x)$ with $N=1$ requires $\nu = 3\nu-1 = \frac{1}{2}$ or $\nu = 1-3\nu = \frac{1}{4}$ if $\hat{N} = 0$ is assumed, or $\nu = \frac{1}{3} - \nu = \frac{1}{6}$ if $\hat{N} = 1$.

For a generic $\hat{u}(x) \equiv u(x)$ given by (4.2), $\hat{\varphi}_x$ does not lie in the same class [specified by (2.3) (2.4)] as $\varphi_x$ unless $\hat{A}_1 = \pm A_1$ and $0 < 1/\nu$ is the even integers $2N+1 \pm (2\hat{N}+1)$, which several MS and Burgers fronts may share, see Fig.6. Otherwise, $\hat{\varphi}_x$ and $\hat{V}$ are not accessible by (2.3) (4.1) if (4.2) holds and $\nu = O(1)$; getting $\hat{V}$ then requires another approach, *e.g.* variational and based on the 'wavefunction' $e^{-\hat{\phi}/2\nu}$ [23], or as $\hat{V} = -\hat{\varphi}_t(\infty, x)$ from a time-wise numerical integration (*e.g.* spectral) of the transient version of (2.2), $\hat{\varphi}_t + \frac{1}{2}(\hat{\varphi}_x)^2 - \nu \hat{\varphi}_{xx} = u(x)$.

Figure 6 shows two curves $\hat{V}$ or $V$ vs. $u_{rms}$ both belonging to $\nu = \frac{1}{2} = 1-\nu = 3\nu-1$ and $\hat{u}(x) = u(x) = \frac{1}{2} A \cos(x) - \frac{1}{4} A^2 \cos(2x)$: one curve is for 'poleless' Burgers *and* MS flames [$\hat{A}_1 = A = -A_1$, $\hat{V} = \frac{1}{4} A^2 = V$], the other is for a MS front with $N=1$ [$A_1 = A$ and $V$ given by (4.3)]. The $\hat{N} = 0$ pattern is the *only* viable Burgers solution if $\nu = \frac{1}{2}$ [$e^{-\hat{\phi}}$ then is the simplest ground-state 'wavefunction' of Whittaker-Hill type [32]]; the $N=0$ or $N=1$ patterns generalize the flat flame [$\varphi(x) \equiv 0$] or the unique nontrivial steady unforced MS solution pertaining to $\nu = \frac{1}{2}$, respectively. The $N=1$ curve has two branches: (i) along the upper one $A>0$, front and forcing function are in phase, and $V \geq \frac{1}{2}$; (ii) a double-valued one where $A < 0$, front and forcing are in phase-opposition, and no solution exists beyond $u_{rms} \approx 0.11$. The turning point does not result from flame structure modifications [33] or too convex curvature effects [34]: similar to sub-wrinkles of parabolic fronts [base-slope $\Phi_x \sim +x$] [29], two branches merge then disappear because of a globally untenable balance between curvature effects, non-linearity, geometrical stretch induced by forcing [but mediated by $\Phi_x = +|A|\sin(x)$] and the DL instability mechanism: to wit, the Burgers front is *not* quenched.

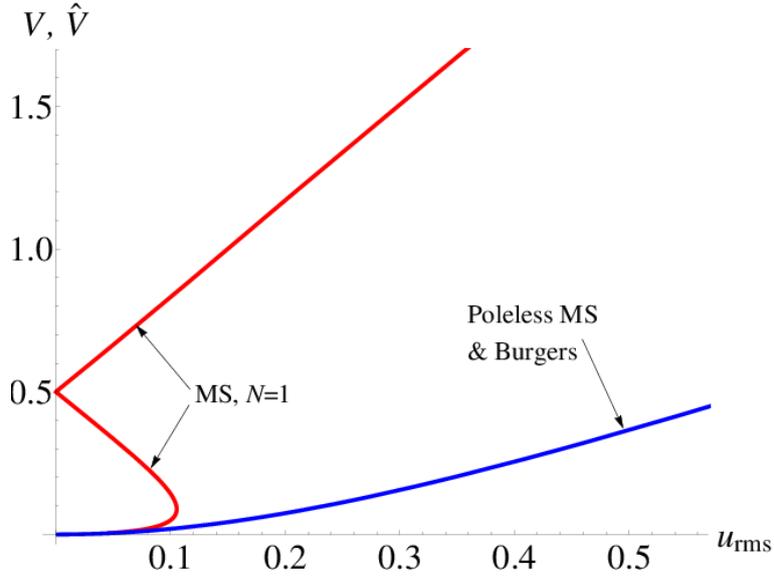

FIG.6.(Color online): Speed increments *vs.* forcing intensity $u_{rms}$ for $\hat{u}(x) = u(x) = \frac{1}{2}A\cos(x) - \frac{1}{4}A^2\cos(2x)$, $\nu = \frac{1}{2}$: the lowest, blue line corresponds to $V = \hat{V}$ for MS *and* Burgers 'poleless' solutions; the other, red lines are $V$ from (4.2) for a MS front with $N = 1$.

## B. Large wrinkles

The matter again simplifies for large MS wrinkles. The material of Sec. **IIIB** can indeed be adapted once the $N \sim 1/\nu \gg 1$ form of (4.2) is envisaged, namely:

$$u(x) = (2\nu N - 1)A_1 \cos(x) - \tfrac{1}{4}A_1^2 \cos(2x) \ . \tag{4.4}$$

The new slope $\phi_x + \Phi_x = \varphi_x = \varphi_{0x}(x) + \varphi_{\star x}(x)$ has the same DL contribution as in (3.14), $\varphi_{0x}(x) = -\tfrac{2}{\pi}\sinh^{-1}(T_{max} / \vartheta(x))$ with $\vartheta(x) = \tan(\tfrac{1}{2}x)$, while $\varphi_{\star x}(x)$ is the double limit of (3.15):

$$\varphi_x(x) = \varphi_{0x}(x) - A_1 \frac{2\vartheta(x)(1 + T_{max}^2 / \vartheta(x)^2)^{1/2}}{(1 - T_{max}^2)^{1/2}(1 + \vartheta(x)^2)} \ . \tag{4.5}$$

Likewise, the density $\rho(B)$ is $\rho_0(B) + \rho_\star(B)$ with $\rho_0(B)$ as in (3.10) and $\rho_\star(B)$ the limit of (3.11). The upper bound $T_{max} = \tanh(\frac{1}{2}B_{max})$ of $\rho(T)$ gets linked to $N$ by the limiting form of the normalization (3.12), while (4.3) gives the speed increment $V$ once $\Sigma_{k=1}^{k=N}\cosh(B_k)$ is replaced by $\int_0^{B_{max}}(\rho_0(B)+\rho_\star(B))\cosh(B)dB$ therein and the integral is done analytically:

$$2\nu N = \tfrac{2}{\pi}\sin^{-1}(T_{max}) + A_1 T_{max}^2/(1-T_{max}^2), \tag{4.6}$$

$$V = \mathcal{V}(\nu N) + \tfrac{1}{4}A_1^2 + \frac{2A_1 T_{max}}{\pi(1-T_{max}^2)^{1/2}} + \frac{A_1^2 T_{max}^2(1-\tfrac{1}{2}T_{max}^2)}{(1-T_{max}^2)^2} \ ; \tag{4.7}$$

alternatively, (4.6)(4.7) follow from the double-limits of (3.12) and, with some care, of (3.20). At fixed $N\nu$, curves $V$ vs. $u_{rms}$ are again obtained in parametric form using $T_{max}$ as parameter: (4.6) gives $A_1$ to be used in $u_{rms}^2 = \tfrac{1}{2}(2\nu N-1)^2 A_1^2 + \tfrac{1}{32}A_1^4$ and in (4.7). The predicted $V$ vs. $u_{rms}$ curves plotted in Fig.7 for sample fixed $N\nu$ s resemble those in Sec.**III**, yet the $N\nu > \tfrac{1}{2}$ ones now extend to $u_{rms} = 0$ as expected from Fig.5 [here $\Theta = 1$, and $\mathcal{V}(n) + 2(n-\tfrac{1}{2})^2 \equiv \tfrac{1}{2}$].

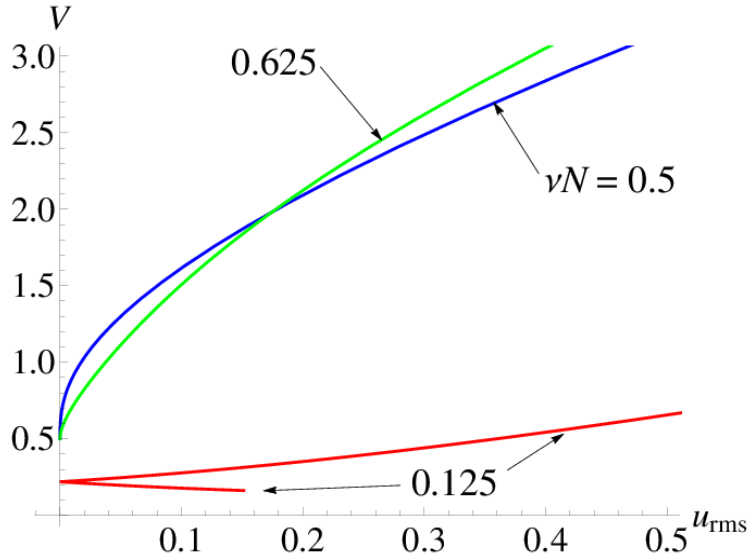

FIG.7. (Color online): Speed increment $V$ vs. forcing intensity $u_{rms}$, plotted parametrically from (4.6)(4.7) with $T_{max}$ as running parameter for three fixed $N\nu$, when the forcing function $u(x)$ reads as in (4.4).

For Burgers fronts also centered on $x=0$, $\hat{\varphi}_x(x)$ reduces to the 'hatted' version $\hat{\varphi}_{\star x}(x)$ of (4.5), and the double limit form of (3.15) gives the density $\hat{\rho}(B) \equiv \hat{\rho}_\star(B)$. The normalization of $\hat{\rho}(B)$ and the ensuing speed increment $\hat{V}$ ultimately read:

$$2\nu\hat{N} = \hat{A}_1\hat{T}_{max}^2/(1-\hat{T}_{max}^2), \tag{4.8}$$

$$\hat{V} = \tfrac{1}{4}\hat{A}_1^2 + 2\nu\hat{N}\hat{A}_1. \tag{4.9}$$

Since $\hat{u}(x) = 2\nu\hat{N}\hat{A}_1\cos(x) - \tfrac{1}{4}\hat{A}_1^2\cos(2x)$ it is readily checked that (4.9) and the counterpart $\hat{\varphi}_{\star x}(x)$ of (4.5) do satisfy the 'easy formula' (3.22) once the normalization (4.8) is employed.

As $u(0) - u(\pi) = 2(2N\nu - 1)A_1$, Burgers fronts belonging to $\hat{u}(x) \equiv u(x)$ and $(2N\nu - 1)A_1 > 0$ have crests centered on $x = 0 (\text{mod}.2\pi)$ by (3.22) [since $u_{min} = u(\pi)$] or by (4.9), or at $x = \pi$ $(\text{mod}.2\pi)$ if $(2N\nu - 1)A_1 < 0$ [$u_{min} = u(0)$]. In either case $\hat{V} = \tfrac{1}{4}A_1^2 + |(2N\nu - 1)A_1|$ is linear in $u_{rms}$ if $u_{rms} \ll 1$ and, with the same slope of $2^{1/2}$, if $u_{rms} \gg 1$; but not quite in between.

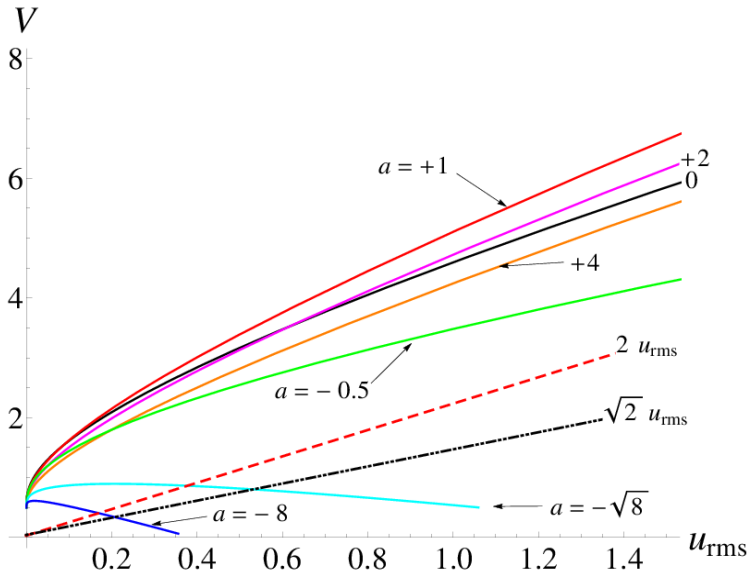

FIG.8. (Color online): MS speed increment $V$ vs. forcing intensity $u_{rms}$ from (4.7)(4.11) (solid lines), for $u(x) = \tfrac{1}{4}A^2(a\cos(x) - \cos(2x))$ and various shape parameters $a$. The straight lines labeled $2u_{rms}$ and $\sqrt{2}u_{rms}$ are the extreme Burgers speed-increments $\hat{V}$ corresponding to $\hat{u}(x) = u(x)$ [i.e. $\hat{V} = u_{rms}[2(1 + |a|)^2/(1 + a^2)]^{1/2}$ at $|a| = 1$ or $|a| + |a^{-1}| \to \infty$, respectively].

As in subsection **III**.B4 one may switch from $(N\nu, A_1)$ to new parameters $(a, A_1)$, where $a$ replaces the former $(\alpha - K)e^\beta$ of (3.24). At fixed $a$, $u(x)$ from (4.4) has a *constant shape*:

$$u(x) = \tfrac{1}{4}A_1^2[a\cos(x) - \cos(2x)], \quad u_{rms} = A_1^2[\tfrac{1}{32}(1 + a^2)]^{1/2}, \tag{4.10}$$

and is separately invariant by $(a, x) \leftrightarrow (-a, x - \pi)$ or $A_1 \leftrightarrow -A_1$. When the new parameters $a := 4(2N\nu - 1)/A_1$ and $A_1$ are given, (4.4)(4.6) [or the double limit of (3.25)(3.24)] imply:

$$\begin{aligned} 2N\nu &= 1 + \tfrac{1}{4}aA_1, \\ [T_{max}^2/(1 - T_{max}^2) - \tfrac{1}{4}a]A_1 &= \tfrac{2}{\pi}\cos^{-1}(T_{max}), \end{aligned} \tag{4.11}$$

hence $\nu N_{u\equiv 0} = \frac{1}{2} = \nu N_{opt}$ at fixed $a$. Using (4.7) this yields $V$ vs. $u_{rms}$ parametrically, see Fig.8.

For *any* $a$, $\cos^{-1}(T_{max} \lesssim 1) \approx (1 - T_{max}^2)^{1/2}$ in (4.11) gives $V - \frac{1}{2} \sim +u_{rms}^{1/3}$ if $u_{rms} \ll 1$, by (4.7).

The $V$ vs. $u_{rms}$ curves belonging to $a > 0$ have a branch where $a/(4+a) < T_{max}^2 \leq 1$, $A_1 > 0$, $2\nu N \geq 1$, and fronts are in phase with $u(x)$; the branch extends to $u_{rms} = \infty$, has $V \sim +u_{rms} + O(u_{rms}^{1/2})$ if $u_{rms} \gg 1$ [see **III**.B3], and follows the scaling law $V - \frac{1}{2} \approx a^{-1} \Upsilon_+(a^2 u_{rms})$ for $a \to +\infty$, with $\Upsilon_+(\upsilon \ll 1) \sim \upsilon^{1/3}$ and $\Upsilon_+(\upsilon \gg 1) \sim \upsilon^{1/2}$. Another branch [not shown] exists if $a > 0$: it has $A_1 < 0$ but is spurious as it violates the double limit $\pi A_1 \geq -(1 - T_{max}^2)^{3/2}$ of (3.13). The curves with $a < 0$ have $0 \leq 2\nu N \leq 1$. They pertain to MS fronts in phase-opposition with $u(x)$, are non-monotonic if $a < -\frac{8}{3}$ and *end* at $B_{max} = 0^+$, $2\nu N = 0^+$, $A_1 = -\frac{4}{a} > 0$, $V = \frac{4}{a^2}$ [$\leq V_{u\equiv 0} = \frac{1}{2}$ if $a \leq -2^{3/2}$]; near the end-point the almost poleless pattern only has small crests at $|x| \sim B_{max} \sim \nu N \ll 1$ and images as localized corrections to $\Phi(x) = A_1 \cos(x)$. Again $u(x)$ gets sinusoidal to leading order if $a \to -\infty$, and another limiting scaling law holds for $0 \leq u_{rms} < \frac{1}{2a^2}(1+a^2)^{1/2} \approx -\frac{1}{2a}$, *viz.*: $V - \frac{1}{2} \approx -a^{-1} \Upsilon_-(a^2 u_{rms})$ with $\Upsilon_-(\upsilon \ll 1) \sim +\upsilon^{1/3}$, $\Upsilon_-(\upsilon \gg 1) \approx -\upsilon$.

Since $u(0) - u(\pi) = \frac{1}{2} a A_1^2$ Burgers flames belonging to $\hat{u}(x) = u(x)$ and $a > 0$ have crests at $x = 0 \pmod{2\pi}$ by (3.22) [$u_{min} = u(\pi)$], or at $x = \pi \pmod{2\pi}$ if $a < 0$ [$u_{min} = u(0)$]; for both, the curve $\hat{V} = -u_{min} = u_{rms}[2(1+|a|)^2/(1+a^2)]^{1/2}$ *vs.* $u_{rms}$ throughout is a plain *straight* line whose slope $2^{1/2} \leq d\hat{V}/du_{rms} \leq 2$ is a decreasing function of $|a| + 1/|a|$ [only].

## V. RICHER BASE-SLOPES

Fairly general periodic base-slopes and associated pole equations are [in obvious notation]:

$$\Phi_x(x) = -\sum_{q=1}^{Q} A_q \sin(qx) , \qquad (5.1)$$

$$\nu \sum_{k \neq j = -N}^{+N} \coth(\tfrac{1}{2}(B_k - B_j)) = \mathrm{sgn}(B_k) + \sum_{q=1}^{Q} A_q \sinh(qB_k) , \qquad (5.2)$$

and similarly for the Burgers fronts. At the expense of a cumbersome algebra, the strategy advocated in Sec. **II** provides formulae for speed increment $V$ and forcing function $u(x)$:

$$V = \mathcal{V}(\nu N) + \tfrac{1}{4} \sum_{q=1}^{Q} A_q^2 + 2\nu \sum_{q=1}^{Q} A_q \sum_{k=1}^{N} \cosh(qB_k) , \qquad (5.3)$$

$$u(x) = \sum_{q=1}^{Q} A_q D_q(x) - \frac{1}{4}\sum_{q=1}^{Q} A_q^2 \cos(2qx) + \frac{1}{2}\sum_{q \neq s} A_q A_s \sin(qx)\sin(sx) . \qquad (5.4)$$

These resume (4.3) and (4.2) if $A_q = 0$ for $q \neq 1$, since $D_1(x) = (2\nu N + \nu - 1)\cos(x)$; the other $D_{q>1}(x)$ obey an inhomogeneous three-term recurrence tied to that of $\cos(qx)$, and read:

$$D_{q>1}(x) = (2\nu N + q\nu - 1)\cos(qx) + 4\nu \sum_{j=1}^{q-1}[\cos(jx)\sum_{k=1}^{N}\cosh((q-j)B_k)] . \qquad (5.5)$$

Burgers fronts have $\mathcal{V}(\nu N)$ replaced by $-2(\hat{N}\nu)^2$ in (5.3) and $q\nu$ instead of $q\nu - 1$ in (5.5).

Given $N$ and $\{A_q\} := A_1,...,A_Q$, (5.2) generically requires numerical iterations to get the $B_k$s; unfortunately, the $D_{q>1}(x)$ and $u(x)$ in (5.5)(5.4) depend on them. Requiring $\hat{u}(x) \equiv u(x)$ will also pose problem, for in general one cannot monitor the $2Q$ Fourier weights of $\hat{u}(x)$ by solely adjusting $\hat{A}_1,...,\hat{A}_Q$ and $\hat{N}$: $\hat{\varphi}_x(x)$ almost never belongs to the same class as $\varphi_x(x)$.

Yet an analysis is still feasible for $N \sim 1/\nu \gg 1$: (3.22) readily gives $\hat{\varphi}_x(x)$, and the MS pole population admits a density $\rho(B,\{A_q\})$ that obeys an equation of same type as (3.7), with $f(B) = f_0(B) + \Sigma_{q \geq 1}(A_q \sinh(qB)/\nu)$ in its right-hand side [RHS]. The linearity of Tricomi equations at fixed $B_{\max}$ implies that $\rho(B,\{A_q\}) \equiv \rho_0(B) + \Sigma_{q \geq 1}A_q \rho_q(B)$, where $\rho_0(B)$ is as in (3.10) and each $\rho_{q \geq 1}(B)$ obeys the integral equation with $\sinh(qB)/\nu$ as RHS. Since $f_\star(B) = m[\coth(\frac{1}{2}(\beta - B)) - \coth(\frac{1}{2}(\beta + B))]$ in (3.7) is $4m\Sigma_{q \geq 1}\sinh(qB)e^{-q\beta}$ for $\beta > B_{\max}$, the known $\rho_\star(T,\beta)$ in (3.11), once expanded as $4m\nu \Sigma_{q \geq 1}\rho_q(B)e^{-q\beta}$, gives *all* the $\rho_{q \geq 1}(B)$s. If expanded as a like power series in $e^{-\beta}$ the rightmost grouping in (3.12) gives $\int_0^{B_{\max}} \rho_j(B)dB := J_{j,0}(T_{\max})$ $= \frac{1}{4\nu}[\mathcal{P}_j(\cosh(B_{\max})) - \mathcal{P}_{j-1}(\cosh(B_{\max}))]$, where the $\mathcal{P}_j(.)$s are Legendre polynomials. The then available normalization of $\rho(B,\{A_q\})$, $N = \frac{1}{\pi\nu}\sin^{-1}(T_{\max}) + \Sigma_{j \geq 1}A_j J_{j,0}(T_{\max})$, provides $T_{\max}$.

The procedure carries over to $\varphi_x - \Phi_x = -\nu \int_{-B_{\max}}^{+B_{\max}} \rho(B,\{A_q\})\cot(\frac{1}{2}(x - iB))dB$. Just like in (3.15) $\Phi_x(x)$ no longer shows up in the 'total' MS slope $\varphi_x(x) = \varphi_{0x}(x) + \Sigma_{q \geq 1}A_q \varphi_{qx}(x)$, where the $\varphi_{qx}(x)$s *all* follow the *known* $\varphi_{\star x}(x,\beta) \equiv 4m\nu \Sigma_{q \geq 1}e^{-q\beta}\varphi_{qx}(x)$ of (3.15). While expressible in terms of $\mathcal{P}_n(\cosh(B_{\max}))$s and trigonometric functions of $x$, *e.g.* using Mathematica [35], the formulae for $\varphi_{qx}(x)$ and $\varphi_x(x)$ get very bulky for $(q,Q) > 3$, though. To illustrate the method, the MS normalization condition and front slope pertaining to $A_{q \neq 3} = 0$, $A_3 \neq 0$ read:

$$2\nu N = \frac{2}{\pi}\sin^{-1}(T_{\max}) + A_3 \frac{T_{\max}^2(T_{\max}^4 + 6T_{\max}^2 + 3)}{(1-T_{\max}^2)^3} \,, \tag{5.6}$$

$$\varphi_x(x) = \varphi_{0x}(x) - A_3 \frac{2\vartheta(x)(1+T_{\max}^2/\vartheta(x)^2)^{1/2}}{(1-T_{\max}^2)^{3-1/2}(1+\vartheta(x)^2)^3}\Pi_3(T_{\max},\vartheta(x)) \,, \tag{5.7}$$

$$\Pi_3(T,\vartheta) := (1-3\vartheta^2)^2 T^4 + (2+28\vartheta^2 - 6\vartheta^4)T^2 + 3 - 10\vartheta^2 + 3\vartheta^4,$$

again with $\vartheta(x) = \tan(\tfrac{1}{2}x)$ and the same DL slope $\varphi_{0x}(x)$ as in (3.14) [compare to (4.5)(4.6)].

The integrals $J_q(T_{\max},\{A_q\}) = \int_0^{B_{\max}} \rho(B,\{A_q\})\cosh(qB)dB = J_{0,q}(T_{\max}) + \Sigma_{j\geq 1} A_j J_{j,q}(T_{\max})$ are also needed as asymptotic estimates of $\Sigma_{k=1}^N \cosh(qB_k)$ to deduce the forcing function $u(x)$ and the speed increment $V$ from (5.3)-(5.5). Appendix **A** sketches how the $J_{j,q}(T_{\max})$ involved are all deducible from a generating function whose structure is inspired by Sec.**III** [see above (3.17)]. For example the integrals $J_q(T_{\max},A_3) = J_{0,q}(T_{\max}) + J_{3,q}(T_{\max})A_3$, $q=1,2,3$, needed if $A_{q\neq 3} = 0$ and $A_3 \neq 0$ as in (5.6)(5.7), are given by Eqs.(A.4)-(A.6) of Appendix **A**, for use in:

$$\begin{aligned} u(x) &= 4A_3[\nu J_2(T_{\max},A_3)\cos(x) + \nu J_1(T_{\max},A_3)\cos(2x)] \\ &\quad + (2N\nu - 1)A_3\cos(3x) - \tfrac{1}{4}A_3^2\cos(6x) \,, \end{aligned} \tag{5.8}$$

$$V = \mathcal{V}(\nu N) + \tfrac{1}{4}A_3^2 + 2A_3\nu J_3(T_{\max},A_3) \,. \tag{5.9}$$

By (3.22), forced Burgers fronts with $\hat{u}(x) = u(x)$ simply have $\hat{V} = -u_{\min}$, $\tfrac{1}{2}\hat{\varphi}_x^{\,2} = u(x) - u_{\min}$ : $\hat{\varphi}(x)$ is sensitive to the small-scale squiggles of $u(x)$ and has two crests if $x_{\min} \neq (0,\pi)$. The additive DL contribution $\varphi_{0x}(x)$ to $\varphi_x(x)$ [*e.g.* in (5.7)] renders MS slopes more robust; this effect likely helps coherent flame wrinkles resist moderate turbulent forcing [Fig.11 in [22]].

Section **V** thus succeeds in giving explicit MS long-waved front slopes, forcing functions and speed increments monitored by $Q+1 > 2$ free parameters $A_1,...,A_Q$ and $N\nu > 0$; $V$ could then be plotted *vs.* $u_{rms}$ *at fixed* $N\nu$, once the $A_q$ are specified. Large $A_q$s lead to a peaked pole-density whose support $T_{\max} \approx \tfrac{1}{2} B_{\max} \ll 1$ obeys a universal normalization where the $O(T_{\max})$ DL contribution again is sub-leading: $2N\nu \approx T_{\max}^2\Sigma_{q\geq 1} qA_q + \tfrac{2}{\pi}T_{\max}$. The $J_q$s featured in the continuous versions of (5.3)-(5.5) indeed *all* simplify to $J_q(T_{\max},\{A\}) \approx N$, whence $V = \tfrac{1}{4}\Sigma_{q\geq 1}A_q^2 + 2N\nu\Sigma_{q\geq 1}A_q + O(1)$; since $u_{rms}^{\,2}$ in (5.4) gets quartic in each of the large $A_1,...,A_Q$, the near-linear law $V \sim u_{rms} + O(u_{rms}^{1/2})$ again ensues. But $u(x)/u_{rms}$ could *not* be put in $u_{rms}$-independent form if $1 < Q < \infty$: being quadratic in $\Phi_x$, $u(x)$ generically comprises too many harmonics for $Q+1$ parameters to control it. Yet some well-chosen full series [$Q = \infty$] can possibly make it: for example, geometric series $A_{q \geq 1} \propto (\pm e^{-\beta})^q$ bring one back to Sec. **III**.

## VI. CONCLUSION

The above work took up the combined influences of curvature effects, Darrieus-Landau [DL] instability, eikonal nonlinearity and external forcing on the shape(s) $\varphi(x)$ and speed(s) $V$ of premixed flames. Steady patterns and forcing functions were envisaged as first step, by means of a steadily forced Michelson-Sivashinsky [MS] equation; by a change of frame, this would also apply to fronts that are phase-locked to travelling shear waves. DL-free fronts and their $\hat{\varphi}(x)$, $\hat{V}$ were also studied for comparison, *via* a time-independent forced Burgers equation.

The proposed semi-inverse method of solution split the MS front slope $\varphi_x$ as $\Phi_x + \phi_x$, where $\Phi_x(x)$ is prescribed and $\phi_x(x)$ is fully defined by its wavelength and its $2N$ poles [per cell]. Equations were derived to locate the poles, in principle giving access to $\varphi_x(x)$, to the flame speed increment $V$ caused by wrinkling and to the forcing function $u(x)$; and similarly for the Burgers analogs. The method was applied to 3 types of base-slopes $\Phi_x$:(i) pairs of cotangents, (ii) sine functions, and (iii) sums of $Q < \infty$ sines. If $O(1)$ neutral-to-actual wrinkle-wavelength ratios $\nu$ are envisaged, numerical iterations are in general needed to locate the poles of $\phi_x(x)$ and deduce $\varphi(x)$, $V$ and $u(x)$; $N = (0,1)$ are notable exceptions for which the curves $V$ *vs.* forcing intensity $u_{rms}$ and their Burgers analogs were obtained analytically (case(ii)), revealing solution multiplicity and a phenomenon of quenching by too strong forcing-induced stretch.

For $N \sim 1/\nu \gg 1$, the poles of $\phi_x$ admit a density, determined in cases (i)-(iii) on solving Tricomi integral equations analytically; the latter's linearity allowed the solutions for case (i) to generate those for (ii)(iii). Closed form speed-increments, front-slope and forcing-function profiles ensued; numerical determinations of the poles of $\phi_x(x)$ confirmed the predictions.

In cases (i)-(iii), MS fronts in phase with $u(x)$ propagate faster than identically forced DL-free fronts; the DL effect can also quench those MS fronts in phase opposition with a too strong forcing. In accord with qualitative estimates all cases gave $\hat{V} \sim u_{rms}$ and $V - \hat{V} \sim +u_{rms}^{1/2}$ at $u_{rms} \gg 1$. The small $u_{rms}$ behaviors differ at fixed pole-number $N$ [obtained in all cases] and at fixed shape $u(x)/u_{rms}$ of $u(x)$ [for (i)(ii) only]. In the latter situation it was found that:

*(a)*: Contrary to Burgers' $\hat{V}$, $V$ has an initial gap $V_{u\equiv 0} \neq 0$; in case (ii), $V_{u\equiv 0} = \frac{1}{2}$ is that of unforced fronts with optimal filling $N = N_{opt}(\nu) \approx \frac{1}{2\nu}$ of the pole pile [see Sec. **I**].

*(b)*: $V(u_{rms}) - V_{u\equiv 0}$ varies roughly as $u_{rms}^{\eta}$, with $\eta(u_{ms})$ slowly decaying from $\eta(0) = \frac{1}{3}$ to $\eta(\infty) = 1$ through a wide $\eta \approx \frac{1}{2}$ transition region, similar to what simulations of "turbulently-forced" MS fronts gave for $\nu \ll 1$ [22]; see also [30, 31]. Similar trends will be obtained whenever the $2\pi$-periodic function $\Phi_x(x+iB)$ is entire and satisfies $i\Phi_x(\pm i\infty) = \pm\infty$.

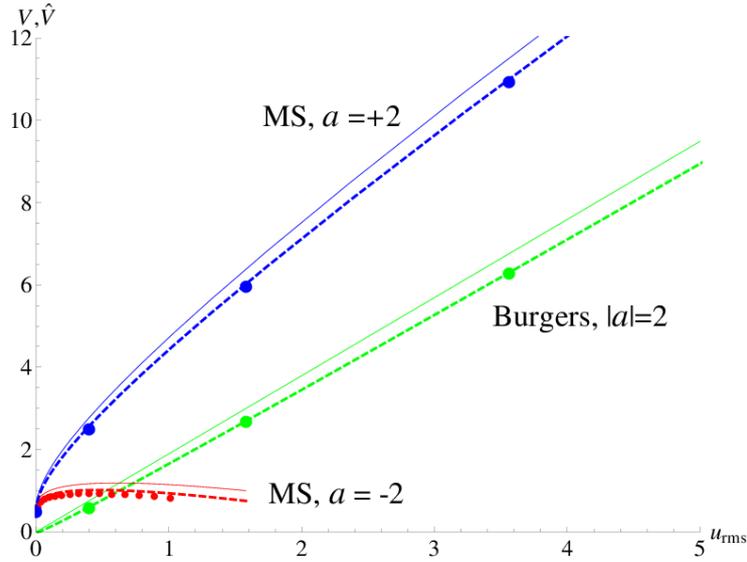

FIG.9.(Color online): Speed increments *vs.* forcing intensity $u_{rms}$ for $\nu = \frac{1}{8}$, $u(x)$ from (4.10): $V$ from (4.7) (4.10)(4.11) for MS flames with $a = 2$ [upper, blue] or $a = -2$ [lowest, red]; upper (blue) or lower (red) dots : $-\varphi_t(\infty, x)$ from the transient MS equation. Middle lines and dots: Burgers $\hat{V}$ and $-\hat{\varphi}_t(\infty, x)$ for $\hat{u}(x) = u(x)$ and $|a| = 2$. Only the dashed lines include curvature effects: $\delta V := -\nu\varphi_{xx}(x_{min}) \approx -\nu(\frac{1}{\pi}T_{max} + A_1(1 - T_{max}^2)^{-1/2})$ from (4.5), and $\delta\hat{V} \approx -\nu(u_{xx}(\hat{x}_{min}))^{1/2}$ from (4.10).

Separate determinations of $V$ as $-\varphi_t(\infty, x)$ from spectral integrations of the transient MS equation $\varphi_t + \frac{1}{2}(\varphi_x)^2 - \nu\varphi_{xx} + \mathcal{H}[\varphi_x] = u(x)$ reveal that pole-densities fairly estimate $V$ even for moderately-long wrinkles, Fig.9. Likewise, for Burgers fronts sharing the same $u(x)$, $\hat{V} \approx -u(x_{min})$ yields a good first-order prediction; and a nearly perfect one results for $\nu \leq \frac{1}{8}$ if one accounts for the curvature correction $\delta\hat{V} = -\nu\varphi_{xx}(x_{min}) \approx -\nu(u_{xx}(\hat{x}_{min}))^{1/2}$, Fig.9. Fortunately, adding the *leading*-order estimate of $\delta V = -\nu\varphi_{xx}(x_{min})$ to the $\nu = 0^+$ MS speed increment also provides excellent predictions of $V$ for all $\nu \leq \frac{1}{8}$, Fig.9, and fair ones up to $\nu = \frac{1}{4}$. To be sure, such curvature-corrected results hold for $v \ll 1$ only outside thin layers in which $u_{rms}$ also is small: at $u_{rms} \sim v^2$ where $\delta \hat{V} \sim \hat{V} \sim v^2$, or at $u_{rms} \sim v^3$ where $\delta V$ and $O(u_{rms}^{1/3})$ contributions to $V - \frac{1}{2} \sim v$ recombine and ensure a smooth transition to the exact $V_{u \equiv 0} = V_0$ recalled in (1.2).

These results carry over to flame *surfaces* involving two Cartesian coordinates $\mathbf{x} = (x_1, x_2)$ in lieu of a single $x$, when the forcing function reads $u(\mathbf{x}) = u_1(x_1) + u_2(x_2)$: the two-dimensional forced version of (1.1) then admits 'separated' solutions $\varphi(t, \mathbf{x}) = -(V_1 + V_2)t + \varphi_1(x_1) + \varphi_2(x_2)$ where the $(V_r, \varphi_r(x_r))$ independently obey (2.1), possibly with $v_1 \ne v_2$; and the same for Burgers fronts. For square cells $[v_1 = v = v_2]$ one has $V_1 = V = V_2$ if $u_1(.) = u_2(.)$, whereby such graphs as in Fig.9 then yields $V = \frac{1}{2}(V_1 + V_2)$ vs. $u_{rms} = u'/2^{1/2}$, with $(u')^2 := u_{1,rms}^2 + u_{2,rms}^2$. As Fig.8 indicates $V \approx 6$ for $u_{rms} = 2^{1/2}$ and $0 \le a \le 2$, the predicted effective burning speed of the wrinkled flame *surface* is $u_L(1 + 2f(E)V) \approx 5.64 u_L$ at $U'/u_L = u' f(E) \approx 0.774$ if $E = 6.5$ [ $f(E) \approx 0.387$ ]; getting such a rough agreement with the turbulent flame speeds in Fig.9 of [31] likely is coincidental, yet the qualitative accord with the present Fig.9 is encouraging.

The above analyses would deserve to be extended. For example, the strategy of Sec. **II** can handle $2N_0$ poles [per cell] at $z = 0 + iB_k$ and $2N_\pi$ ones at $z = \pi + ib_s$. If $N_0 \sim N_\pi \sim 1/v \gg 1$ the *two* pole-densities belonging to such bi-coalesced [or two-crested] fronts obey a pair of coupled integral equations that are easy to write [15, 36] but are still awaiting solutions.

Retaining more orders of the expansion in Atwood-number $\mathcal{A}$ will improve accuracy and ease comparisons with experiments, but a new nonlinearity $\frac{1}{2}c(\mathcal{A})(\mathcal{H}[\varphi_x])^2 < 0$ first appears on the left of (2.1) [37] ; the strategy of Sec. **II** can handle it, yet solving the enlarged steady forced equation to get $\phi_x(x + iB)$ for $N \sim 1/v \gg 1$ entails adaptation of the complex-variable techniques used in [36]. Although cotangent base-slopes [Sec.**III**] and their sinusoidal limit [Sec.**IV**] seem within reach, and the key property of linearity [Secs. **III**, **V**] still holds, the analysis needed to get the $c(\mathcal{A})$-modified $V, \varphi_x(x)$ and $u(x)$ is not available to date.

Exactly solving (2.1) with any *prescribed* $u(x)$ [*direct* problem] is hopeless: this already is the case for (2.2) in Schrödinger form. Yet, instead of explicitly given $\Phi_x$ could be defined by an equation, *e.g.* $\frac{1}{2}(\Phi_x)^2 - \mu \Phi_{xx} + \lambda \mathcal{H}[\Phi_x] = W + w(x)$ for constant $(\mu, \lambda, W)$ and a $w(x) = w(x + 2\pi)$; this can hopefully help one extend Darboux's trick [23]. Identifying all $\Phi_x$s that lead to $u_{rms}$-independent $u(x)/u_{rms}$ functions would also be rewarding: for the mathematics of flame theory, and because works of the type initiated here can yield values $V_{av}$ of $V$ that are *ensemble*-averaged at fixed $u_{rms}$ over the parameter(s) monitoring the shape of $u(x)$ [*e.g.*, $a$ in Sec. **IV**], which is not so different from what some experiments deliver [31]; $V_{av}(u_{rms})$ also feels the fronts in phase-opposition with $u(x)$, which will somewhat soften its initial growth.

But *the* real theoretical problem to address next concerns unsteadily-forced flames. Beside the analytical challenge, access to *time*-averaged-$V$ *vs.* $u_{rms}$ curves could be of interest for large-eddy simulations of turbulent flames if the smallest computational grid cells, say squares of side $\delta$, still contain a piece of *wrinkled* front: the flow-field disturbances that cause the DL instability have an upstream range $\leq \delta/2^{1/2}\pi$, whence a sub-grid analysis is needed to get the DL-affected effective flame speed $u_e$ belonging to such grid cells. Unsteadily forced flames at least require tackling unsteady pole equations, however, and will be reported on elsewhere.

## ACKNOWLEDGEMENTS

GJ thanks H. El-Rabii (CNRS and Poitiers University) for discussions and help as to figures. BD's work has been carried out in the framework of Labex MEC (ANR-10-LABX-0092) and of A*MIDEX project (ANR-11-IDEX-0001-02), funded by the ''Investissements d'Avenir'' French Government program managed by the French National Research Agency (ANR).

## APPENDIX A: INTEGRAL-GENERATING FUNCTION

The integrals $J_{j,q}(T_{\max}) := \int_0^{B_{\max}} \rho_j(B)\cosh(qB)dB$ result from identifying two evaluations of :

$$G(\beta,\sigma) := \int_{-B_{\max}}^{B_{\max}} \rho(B,\beta)\coth(\tfrac{1}{2}(\sigma+B))dB, \qquad (A.1)$$

where $(\beta,\sigma) > B_{\max}$ and $\rho(T,\beta) := \rho(B,\beta)|_{B=2\tanh^{-1}(T)}$ is $\rho_0(T) + \rho_\star(T,\beta)$ as in (3.10)(3.11).

Analytical evaluation of the integral in (A.1), using (3.10)(3.11), first yields:

$$G(\beta,\sigma) = \frac{2}{\nu\pi}\sin^{-1}(\frac{T_{\max}}{\Sigma}) + \frac{2m(1-\Theta^2)}{(1-T_{\max}^2/\Theta^2)^{1/2}} \\ \times \frac{\Sigma}{\Theta^2-\Sigma^2}[(1-\frac{T_{\max}^2}{\Theta^2})^{1/2} - (1-\frac{T_{\max}^2}{\Sigma^2})^{1/2}], \qquad (A.2)$$

with $\Theta = \tanh(\tfrac{1}{2}\beta) = (1-e^{-\beta})/(1+e^{-\beta})$, $\Sigma = \tanh(\tfrac{1}{2}\sigma)$. Next $\coth(\tfrac{1}{2}(\sigma+B)) + \coth(\tfrac{1}{2}(\sigma-B))$ $\equiv 2 + 4\Sigma_{q=1}^{q=\infty}e^{-q\sigma}\cosh(qB)$ and $\rho(T,\beta) \equiv \rho_0(B) + 4m\Sigma_{j=1}^{j=\infty}e^{-j\beta}\rho_j(B)$, whereby (A.1) also is:

$$G(\beta,\sigma) = 2J_{0,0} + 4\sum_{q=1}^{\infty} e^{-q\sigma} J_{0,q} + 8m\sum_{j=1}^{\infty} e^{-j\beta}[J_{j,0} + 2\sum_{q=1}^{\infty} e^{-q\sigma} J_{j,q}] \ . \tag{A.3}$$

Identifying (A.3) and (A.2) [expanded as a double series in $e^{-j\beta}e^{-q\sigma}$] gives all the $J_{j,q}(T_{\max})$ s. The $J_q(T_{\max},\{A_q\})$ evoked below (5.7) read as $J_{0,q}(T_{\max}) + \Sigma_{j\geq 1} A_j J_{j,q}(T_{\max})$. In particular, those integrals needed in (5.8)(5.9), $J_q(T_{\max},A_3) = J_{0,q}(T_{\max}) + J_{3,q}(T_{\max})A_3$, are:

$$\nu J_1(T_{\max},A_3) = \frac{T_{\max}}{\pi(1-T_{\max}^2)^{1/2}} + A_3 \frac{3T_{\max}^2(3T_{\max}^2+2)}{4(1-T_{\max}^2)^4} \ , \tag{A.4}$$

$$\nu J_2(T_{\max},A_3) = \frac{T_{\max}}{\pi(1-T_{\max}^2)^{3/2}} + A_3 \frac{3T_{\max}^2(3T_{\max}^4+2T_{\max}^2+1)}{2(1-T_{\max}^2)^5} , \tag{A.5}$$

$$\nu J_3(T_{\max},A_3) = \frac{T_{\max}(T_{\max}^4+2T_{\max}^2+3)}{3\pi(1-T_{\max}^2)^{5/2}} + $$
$$A_3 \frac{T_{\max}^2(-T_{\max}^{10}+6T_{\max}^8+12T_{\max}^6+56T_{\max}^4+21T_{\max}^2+6)}{4(1-T_{\max}^2)^6} \ . \tag{A.6}$$

## **REFERENCES**.

---------------------------